# Unraveling Structural and Magnetic Information during Growth of Nanocrystalline SrFe$_{12}$O$_{19}$

Cecilia Granados-Miralles[1], Matilde Saura-Múzquiz[1], Espen D. Bøjesen[1], Kirsten M. Ø. Jensen[2],

Henrik L. Andersen[1] & Mogens Christensen*[1]

1 : Center for Materials Crystallography, Department of Chemistry and iNano, Aarhus University,
Langelandsgade 140
DK-8000 Aarhus C, Denmark

2 : Nanoscience Center, Department of Chemistry, University of Copenhagen,
Universitetsparken 5
DK-2100 København Ø, Denmark

* corresponding Author:
Mogens Christensen
mch@chem.au.dk

# Abstract

The hydrothermal synthesis of magnetic strontium hexaferrite ($SrFe_{12}O_{19}$) nanocrystallites was followed *in situ* using synchrotron powder X-ray diffraction. For all the studied temperatures, the formation of $SrFe_{12}O_{19}$ happened through an intermediate crystalline phase, identified as the so-called six-line ferrihydrite (FeOOH). The presence of FeOOH has been overlooked in previous studies on hydrothermally synthesized $SrFe_{12}O_{19}$, despite the phase having a non-trivial influence on the magnetic properties of the final material. The chemical synthesis was successfully reproduced *ex situ* in a custom-designed batch-type reactor that resembles the experimental conditions of the *in situ* setup, while allowing larger quantities of material to be produced. The agreement in phase composition between the two studies reveals comparability between both experimental setups. Hexagonal platelet morphology is confirmed for $SrFe_{12}O_{19}$ combining Rietveld refinements of powder X-ray diffraction (PXRD) data with transmission electron microscopy (TEM). Room temperature magnetization curves were measured on the nanopowders prepared *ex situ*. The magnetic properties are discussed in the context of the influence of phase composition and crystallite size.

# 1. Introduction

Rare-earth (RE) magnets are made from metallic alloys that contain elements from the lanthanide series, also known as RE elements. Their high magnetic moments at the atomic level, originating from the unpaired electrons populating the strongly localized f-orbitals, combined with their high magnetocrystalline anisotropy, which translates into a high coercivity ($H_c$), make them the best-performing known permanent magnets (PMs).[1] Nonetheless, RE-based magnets became highly critical materials in 2011 when China, responsible for more than 95% of the world's RE production, introduced restrictions on the exports.[2] Consequently, the search for RE-free alternatives has become a key research topic. Nanostructuring as a means of achieving a higher $H_c$ is among the investigated approaches, and it is addressed in the present work.

As observed for other physical properties, the laws governing magnetism at the nanoscale differ from those of the bulk.[3] Magnetic nanoparticles with a size below a certain value consist of a single magnetic domain, which favors a larger $H_c$ value compared to that of the multi-domain state exhibited by a bulk magnetic material. On the other hand, if the size is too small, the magnetic nanoparticles lose their effective coercivity as they reach the so-called superparamagnetic state. The size range within which a nanoparticle has a single magnetic domain without presenting superparamagnetic behavior is called the stable single-domain (SSD) region and it essentially depends on the intrinsic magnetocrystalline anisotropy of the specific system.[4,5]

Magnetic nanoparticles are the basis of both well-established and developing applications, such as ultrahigh-density data storage media,[6–9] magnetocaloric refrigeration,[10] magnetically-guided drug delivery,[11] localized heating of cancerous cells (hyperthermia),[12–14] or magnetic resonance imaging (MRI).[15] Determining the size-range in which a particular compound behaves as a single magnetic domain is not only relevant to maximize $H_c$, but it is also a crucial parameter for many of the applications listed. For example, in the field of magnetic recording media, high-density storage can be achieved by reducing the size of the magnetic bit, but in order to keep the data usable the superparamagnetic relaxation of the magnetization direction must be avoided.[16] Recent advances in micromagnetic calculations allow a rather accurate estimation of the SSD region.[17,18] Unfortunately, achieving a realistic computational description for the polydisperse outcome of a large scale synthesis is not always straightforward. Some of the crucial parameters, such as particle size and shape, lattice defects or crystallinity, are not well represented by single values. Instead, they are better described as distributions, which are in general difficult to predict with accuracy. Therefore, an experimental determination of the SSD becomes imperative.

Among the RE-free hard magnetic phases, $SrFe_{12}O_{19}$ stands out as a very good candidate, owing to its low cost,[1] non-toxicity,[19] chemical stability, elevated Curie temperature and relatively high anisotropy constant (i.e., high coercivity).[20] Several studies have shown that a top-down preparation of nanosized $SrFe_{12}O_{19}$ by means of, e.g., ball-milling decreases the crystallinity and leads to formation of impurity phases, which in turn diminishes the magnetic properties.[21–23] Therefore, bottom-up chemical routes are preferred. The method of choice in the present study was a hydrothermal synthesis under near-critical conditions, known to be a quick, green and up-scalable method for producing nanoparticles.[24] Furthermore, the characteristic plate-like morphology of the $SrFe_{12}O_{19}$ nanoparticles obtained through this method has been shown to be ideal for achieving a high-degree of alignment in the subsequent compaction process required to transform the nanopowders into a dense magnet.[25,26]

When high control of the properties of the final product is desired, and size being such a crucial parameter for the magnetic properties, an in-depth understanding of the preparation route is paramount. A rich insight on the chemical synthesis can be obtained from studying the process *in situ*. However, hydrothermal syntheses are usually carried out inside sealed thick-walled reaction vessels, due to the relatively high pressures and temperatures, which make these studies experimentally challenging. This has been an ongoing concern during the last decades, and several groups have worked on designing reactors suited for *in situ* scattering studies of these processes.[27–29] A good example is the laboratory-sized synthesis vessels (*ca.* 5-20 mL) designed by O'Hare *et al*.[30–33] A wide variety of inorganic reactions have been followed with a time resolution down to 35 seconds, mostly using energy-dispersive X-ray diffraction (EDXRD).[34–37] Becker *et al.* developed an alternative setup that permits measuring angular-dispersive X-ray diffraction, increasing the Q-coverage and peak resolution of the diffraction data compared with the energy-dispersive mode.[38] Furthermore, the reactor requires very small sample volumes (*ca.* 0.02 mL), which allows reducing the exposure time to below 5 seconds, while still collecting diffraction data of sufficient quality for detailed Rietveld refinements. The setup has been shown to provide very useful knowledge about the crystallization mechanisms of numerous functional inorganic materials,[39,40] including magnetic systems.[41–43] Unfortunately, in most of these studies the physical properties of the product could not be measured, due to the small volume of the *in situ* setup. In addition, it was not always possible to fully transfer the knowledge to a real laboratory synthesis, since conventional autoclaves with their large steel body lead to much slower heating rates. Consequently, we have now designed a laboratory-scale batch-type reactor that matches the reaction conditions of the aforementioned *in situ* setup (by Becker *et al.*), yet allowing *ca*. 100 times larger reaction volumes (*ca.* 2 mL). The product obtained from the 2 mL precursor is sufficient for characterization of material properties.

In the present work, single-phase $SrFe_{12}O_{19}$ was synthesized by a hydrothermal method using the two reactors described above. First, the chemical reaction was followed *in situ* using the setup reported by Becker *et al*. Afterwards, the same synthesis conditions were reproduced in the laboratory (*ex situ*), where larger quantities of sample were prepared, allowing characterization of the magnetic properties of the obtained nanopowders as well as high-resolution powder X-ray diffraction (PXRD) and transmission electron microscopy (TEM).

## 2. Experimental Section

### 2.1. Preparation of the precursor gel

A 2.00 M solution of $Fe(NO_3)_3 \cdot 9H_2O$ was prepared dissolving the chemical in deionized water. 2.00 M $Sr(NO_3)_2$ and 16.00 M NaOH were likewise prepared. All used chemicals were reagent grade from Sigma-Aldrich (purity ≥ 98%).

First, 2.00 mL of the Fe nitrate solution were mixed with 0.25 mL of the Sr nitrate solution. These volumes match the Fe/Sr molar ratio of 8 previously reported to be the optimal to obtain the pristine product by a similar synthesis approach.[44] Subsequently, 1.00 mL of 16.00 M NaOH was added dropwise to the nitrate-containing solution under constant magnetic stirring, causing the transparent solution to become a gel. The added amount of base [OH]⁻ represents an excess of 25% compared to the molar amount of nitrates [NO₃]⁻ present in the solution, thereby ensuring total precipitation of the metallic species. Finally, 2.75mL of deionized water were added to the mixture, leading to a final metal concentration of 0.75 M. After approx. 10 min of magnetic stirring, the so-prepared precursor gel was introduced into the reactor, where upon subsequent pressurization and heating the $SrFe_{12}O_{19}$ crystallization occurred. An identically prepared precursor was used for both *in situ* and *ex situ* experiments.

## 2.2. *In situ* powder X-ray diffraction experiments

### Experimental setup: *in situ* reactor

The chemical synthesis was conducted in a previously described apparatus designed for *in situ* monitoring of hydrothermal reactions using PXRD.[38] The reactor consists of a single-crystal sapphire tube (sample volume ≈ 0.02 mL) in which the precursor gel was injected and pressurized to 250 bars. The synthesis was initiated by a very rapid heating using a preheated hot-air gun placed underneath the tube. Further experimental details, including temperature calibration, are described in the Electronic Supplementary Information (ESI). The whole synthesis process was monitored using synchrotron radiation with a wavelength of 0.9918 Å at the beamline I711, MAX II Synchrotron (MAX-lab, Sweden).[45] PXRD data were collected in 5 second intervals with a large area detector (Titan CCD, ⌀ = 165 mm, Agilent Technologies) until the reaction was completed.

### Powder X-ray diffraction data analysis: Rietveld refinements

The diffraction data were collected as two-dimensional images, which were integrated to one-dimensional 2θ-scans using the software *Fit2D*.[46] Sequential Rietveld refinement of each dataset was performed using *FullProf Suite*.[47] Further details on both data integration and sequential Rietveld refinements may be found in the ESI. The crystallographic model used to describe the data considers two different phases, with crystalline structures shown in Figure 1. The structure of $SrFe_{12}O_{19}$ belongs to the hexagonal space group $P6_3/mmc$ (194) with unit cell parameters *a* = 5.8844(6) Å and *c* = 23.050(3) Å at room temperature.[48] However, the structure of FeOOH is not as well-known. This ferric oxyhydroxide or ferrihydrite is a very common chemical form of $Fe^{3+}$ in nature, where it is found in mineral deposits of a wide variety of near-surface environments. It is often referred to as the six-line phase, as a consequence of the six strong diffraction peaks in the powder pattern using Cu radiation. Both when found in nature and when synthetically produced in a laboratory, it generally occurs as a highly defective and nano-sized material. The scarcity of long-range repetitive units has hampered an accurate structure determination, despite the use of diverse techniques, such as extended X-ray absorption fine structure (EXAFS),[49] PXRD,[50] neutron powder diffraction (NPD),[51] or pair distribution function (PDF) analysis.[52] The precise crystal structure is still a matter of controversy and there is no widely accepted chemical formula. In this work, the phase was described using the hexagonal space group P-31c (163) with unit cell dimensions $a$ = 2.9374(2) Å and $c$ = 9.303(2) Å. This description bears resemblance to the defect-free structure published by Jansen *et al.* in 2002,[51] but both the position and occupancy of the iron atom were refined. The six-line phase was refined as $Fe_{0.86}OO$, ignoring potential oxygen vacancies and the deviation from charge balance. Hydrogen atoms were not included in the model. Detailed information on atomic positions and site occupation is presented in the ESI.

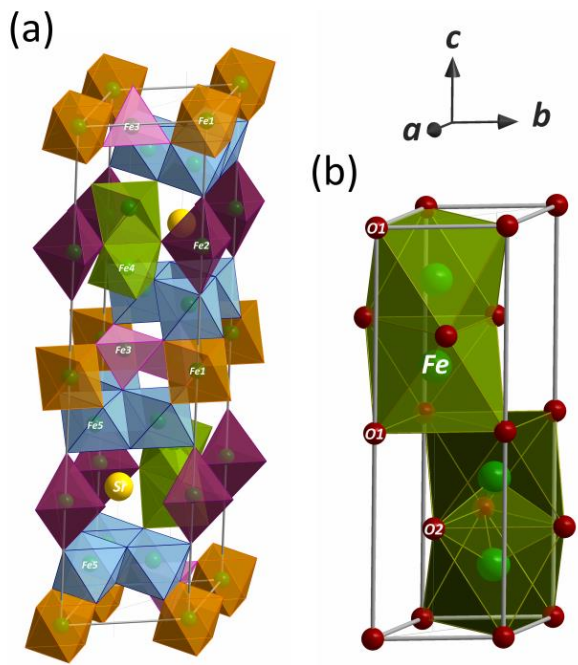

Figure 1. (a) Crystal structure of $SrFe_{12}O_{19}$. $Sr^{2+}$ are represented in yellow and $Fe^{3+}$ in green. For clarity, oxygen atoms are not displayed. Fe atoms occupy five different crystallographic sites, namely octahedral on the corners and edges (orange), trigonal bipyramidal (dark pink), tetrahedral (light pink), face-sharing (green), and edge-sharing (blue) octahedral. (b) Crystal structure of the six-line phase FeOOH, here $Fe^{3+}$ are represented in green and $O^{2-}$ in red. The Fe-sites are *ca.* 40% occupied. Hydrogen atoms were not included in the model.

In the Rietveld refinements, the background of the PXRD patterns was described with a Chebyshev polynomial of 11 refinable coefficients, which provided a good description of the large background from water. A Thompson-Cox-Hastings pseudo-Voigt function convoluted with axial divergence asymmetry was used to model the peak profile.[53] In order to extract the instrumental contribution to the peak broadening, a NIST $LaB_6$ 660b standard[54] was measured in the same experimental configuration as the samples, and refined. The instrumental contribution was deconvoluted from the experimental data, ensuring that the calculated Rietveld model only describes the sample-specific contributions. All sources of peak broadening other than size were neglected. Both the $SrFe_{12}O_{19}$ and FeOOH crystallites were modeled as platelets with large dimensions along the *a/b*-crystallographic plane, *i.e.*, normal to the [001] crystallographic direction. Both isotropic (Y) and anisotropic ($S_z$) contributions to the Lorentzian size were refined. Time-resolved values for weight fractions and crystallite sizes of the two phases were obtained for each dataset.

## 2.3. *Ex situ* laboratory syntheses

### Experimental setup: spiral reactor

A new laboratory setup, in the following referred to as spiral reactor, was specially designed for the purpose of preparing larger quantities of sample under similar conditions to those of the *in situ* experiments. It is a batch-type reactor but it offers a much faster heating rate than conventional autoclaves. The reactor itself consists of approx. 3.5 m Swagelok© stainless steel tubing coiled into a spiral shape. The outer diameter of the tubing is 1/16" (1.59 mm) and the wall thickness is 0.014" (0.88 mm). The reactor can hold approximately 2 mL of precursor gel. A schematic illustration of the experimental setup is shown in Figure 2.

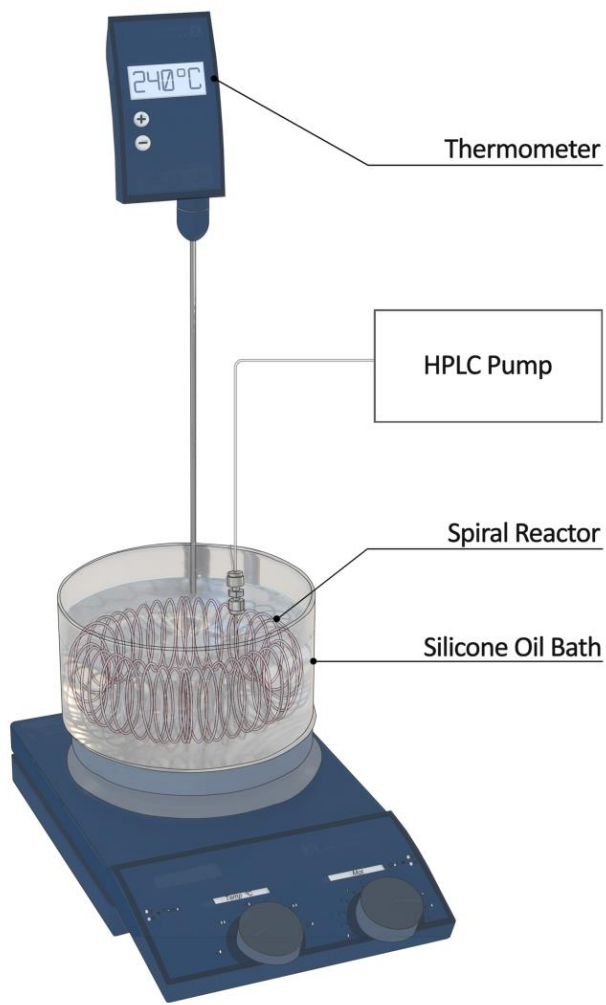

Figure 2. Schematic illustration of the spiral reactor used for the *ex situ* syntheses. The spiral-shaped stainless steel tube is filled with precursor, pressurized with an HPLC pump, and submerged into a preheated silicone oil bath to initiate the $SrFe_{12}O_{19}$ crystallization. Figure made by Clara L. Martín Águila.

The precursor gel was injected into the spiral-shaped tubing and this was sealed with a stainless steel plug at one end. The other end of the spiral was attached to an HPLC pump (Flash 100 Pump, Scientific Systems) pressurizing the system to 250 bar. The filled and pressurized reactor was then immersed into a hot silicone oil bath to initiate the particle formation. The temperature of the silicone oil bath was kept stable by use of a hot plate with temperature feedback and by ensuring a relatively large volume of oil (approx. 1 L) to provide for a large reservoir of heat. The working temperature was always below 280 °C - the boiling point of the oil. The red line in Figure 3 shows the heating curve measured with a K-type thermocouple inside the reactor when fully submerged in silicone oil at 240 °C. It is noteworthy that the set temperature is reached after approximately 20 seconds. For all syntheses the reaction was rapidly quenched by submerging the spiral into a water bath after varying reaction times. From the cooling curve shown in Figure 3 (blue line) it can be seen that the samples reached room-temperature upon quenching within *ca.* 10 seconds.

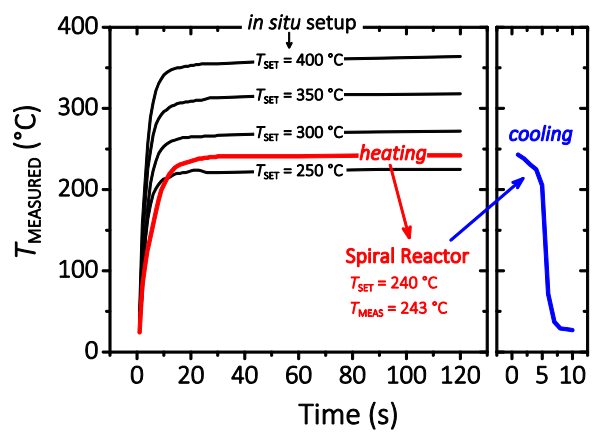

Figure 3. Heating curves for the two experimental setups measured with a K-type thermocouple. The red line shows the heating curve of the spiral reactor; a measured temperature of 243 °C was reached for a setpoint of 240 °C. The cooling rate for the spiral setup is shown by the blue line. For the *in situ* setup, the heating curves at various setpoints are represented in black.

The narrow cross section and the thin walls of the tubing of this reactor resulted in a very efficient heating (and cooling), while still allowing a reasonably large amount of product (*ca.* 100 mg per batch). Furthermore, the spiral reactor matched the heating rates of the *in situ* setup well. The black curves in Figure 3 show the temperature measured at various setpoints for the *in situ* setup.

A time-series of experiments was performed in the spiral reactor at 243 °C, intending to reproduce the *in situ* experiment carried out at 245 °C. The duration of the experiments was 20 s, 40 s, 1 min, 2 min, 5 min, 10 min, and 20 min. The reaction product was collected as a suspension of nanoparticles in water. This suspension was firstly centrifuged (3000 rpm) and washed with 4 M $HNO_3$, then two times with water, and finally with ethanol. This washing process is meant to eliminate non-reacted salts and dissolve potential impurities, such as $SrCO_3$, which often appears due to exposure of unreacted Sr in solution to air.[44] Three reaction batches were produced for each reaction time, leading to *ca.* 300 mg of dry nanopowders.

### High-resolution powder X-ray diffraction

Acquisition of high-resolution PXRD data was possible for the *ex situ* samples. The data were collected on a laboratory diffractometer (Rigaku SmartLab® Diffractometer) using Cu K$_{\alpha1,2}$ radiation. The instrument was operated at 40 kV and 180 mV, in parallel beam configuration and with a D/teX Ultra detector in fluorescence suppression mode. Rietveld analysis of these data was performed following essentially the same procedure described above for the *in situ* synchrotron data. The only difference was the description of the background for the short reaction time samples (20 s, 40 s and 1 min). More details may be found in the ESI.

### Transmission electron microscopy

Micrographs of selected *ex situ* synthesized samples were collected using a TALOS F200A with a TWIN lens system, an X-FEG electron source, and a Ceta 16M camera. A small amount of nanoparticles was suspended in ethanol using an ultrasonic bath (approx. 30 min). A few drops of this suspension were later deposited onto a TEM copper grid coated with lacey carbon.

### Magnetic hysteresis at room temperature

Magnetization at 300 K was measured using a Vibrating Sample Magnetometer (VSM option for the Physical Property Measurement System PPMS®, Quantum Design) varying the applied field $H_{app}$ in the range ±20 kOe. Prior to the magnetic measurements, the powders were dried at 60 °C in a vacuum furnace in order to eliminate possible traces of moisture. Immediately after, the powders were gently compressed using a handheld pressing tool to avoid motion of the particles during the measurement. The measured sample was a cylindrical pellet of 5-10 mg, with a diameter of 3 mm and a thickness of ≈ 1 mm.

Several parameters are extracted from the magnetic hysteresis curve, such as the maximum possible magnetization state of the sample for a specific applied field (saturation magnetization, $M_s$), the residual magnetization remaining in the sample upon removal of the applied field (remanence, $M_r$), as well as the external magnetic field required to completely demagnetize the sample (coercivity, $H_c$). The magnetic energy stored per unit volume of sample (maximum energy product, $BH_{max}$) can be assessed from the hysteresis loop. Further details on $BH_{max}$ calculation are described in the ESI.

# 3. Results

## 3.1. *In situ* study

The hydrothermal synthesis of $SrFe_{12}O_{19}$ was investigated *in situ* at temperatures of 245, 264, 292 and 338 °C, corresponding to temperature setpoints of 270, 290, 320 and 370 °C, respectively. Powder X-ray diffraction data were collected for 20 min with a time resolution of 5 seconds. The two 3-D plots in Figure 4 show the time-evolution of the diffraction data collected at 245 °C.

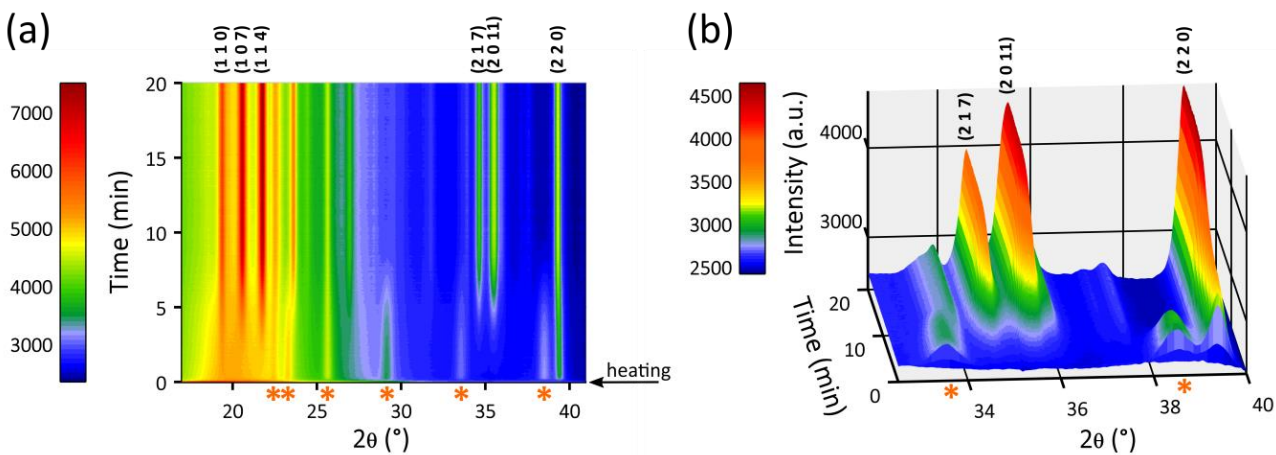

Figure 4. Dataset collected for the *in situ* experiment performed at 245°C, with heating starting at time = 0min. PXRD patterns are plotted vs. reaction time, with a time resolution of 5 seconds, and using a color scale for the intensity. (a) Wide angular range seen from above, *i.e.*, along the intensity axis. (b) Narrower 2θ-range shown in perspective. The orange stars at the bottom of the graph indicate the diffraction peaks corresponding to the six-line FeOOH. The $SrFe_{12}O_{19}$ (*hkl*) reflections referred to in the text are labeled in black at the top of the figure.

The initial PXRD pattern, collected during the first 5 seconds of heating, shows a big hump at around 2θ = 20° and a smaller one at 2θ = 38°, but no Bragg peaks are discernible, highlighting the amorphous nature of the precursor. After the first 5 seconds, peaks corresponding to the six-line FeOOH appear. After approximately 1 min of reaction, new reflections arise; these correspond to the $SrFe_{12}O_{19}$ structure. In addition, it is also observed that not all $SrFe_{12}O_{19}$ peaks appear simultaneously. For example, the two reflections indexed as (110) and (220) emerge almost immediately, while reflections with a large $c$ contribution take longer time to appear, *e.g.*, (107), (217) and (2011). This suggests that initially, very thin platelets parallel to the $a/b$-crystallographic plane are formed. Subsequently, the succeeding crystal growth mainly occurs by increasing the thickness of the platelets, *i.e.*, along the $c$-axis, while less growth is observed along the $a$- and $b$-axes. After about 5 min of reaction at 245 °C, the FeOOH peaks start to disappear, while those of the $SrFe_{12}O_{19}$ become more intense. After 10 min the only observable phase is $SrFe_{12}O_{19}$. This phase transition happens for all four temperatures studied, albeit at different rates.

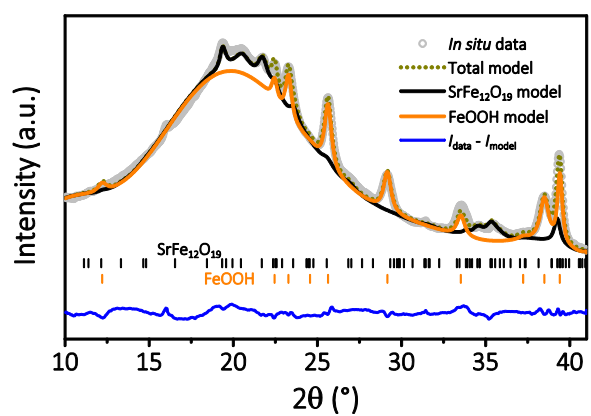

Figure 5. *In situ* PXRD experimental data after 2 min of reaction time for the synthesis performed at 245 °C, along with the corresponding Rietveld model. The measured data are represented by the grey circles, while the models for each of the phases are solid lines: black for $SrFe_{12}O_{19}$ and orange for FeOOH. The total model is represented by the green dotted line. The blue line at the bottom is the difference between the measured data and the calculated Rietveld model.

Sequential Rietveld analysis of the datasets collected at the different temperatures allowed extraction of quantitative information about the chemical processes occurring. As an example, Figure 5 shows the measured and Rietveld modelled diffraction patterns after a reaction time of 2 min for the experiment conducted at 245 °C.

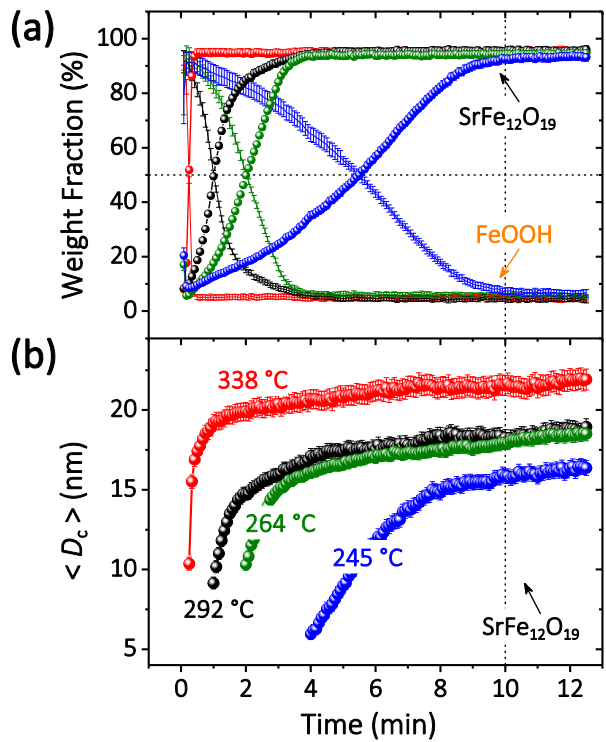

Figure 6. Time-resolved parameters obtained from sequential Rietveld refinement of four *in situ* datasets. Each temperature is represented in a different color: 245 °C in blue, 264 °C in green, 292 °C in black and 338 °C in red. (a) Phase composition of the sample: $SrFe_{12}O_{19}$ (closed spheres), FeOOH (vertical sticks). (b) Thickness of the SFO platelets: crystallite size along the *c*-axis. For clarity, only the first 12 min of data are shown in this figure, although the reaction was followed for 20 min.

Weight fractions of the two phases obtained from Rietveld refinement are shown in Figure 6(a). For all four temperatures under study, the initial FeOOH is completely transformed into $SrFe_{12}O_{19}$. The higher the temperature, the faster the phase transition occurs: it is almost instantaneous for the highest temperature (338 °C), while it takes about 10 min for the lowest (245 °C). A time of around 5 min appears sufficient for the full transformation in the case of the two intermediate temperatures, 264 and 292 °C. Furthermore, it may be noticed that for all temperatures the weight fractions of the two phases are symmetric with respect to an axis located at a weight fraction = 50%. This suggests that the $SrFe_{12}O_{19}$ is originated from the initially formed FeOOH. This phase transformation is very likely a dissolution-recrystallization process, where the FeOOH that crystallizes initially is later re-dissolved and starts to incorporate Sr in the structure, to finally evolve towards $SrFe_{12}O_{19}$.

Both $SrFe_{12}O_{19}$ and FeOOH form crystallites with an anisotropic plate-like shape. The thickness of the platelets for both phases is given by the size along the *c*-direction $<D_c>$, while the diameter corresponds to the size on the *a/b*-plane. Volume-averaged thickness $<D_c>$ of the $SrFe_{12}O_{19}$ platelets is displayed in Figure 6(b). The platelets diameters are not shown since they are beyond the resolution limit of the instrumental configuration. Crystallite thickness increases with reaction time, although it seems to do so only up to a certain maximum value determined by the reaction temperature: the higher the temperature, the larger the final crystallite thickness, ranging between 17 and 22 nm. The crystallite growth also depends on the reaction temperature, and it occurs faster at higher temperatures.

In order to improve the understanding of the reaction, the time-resolved data were fitted to the Johnson–Mehl–Avrami–Kolmogorov (JMAK) kinetic model. Although this model was originally developed for solidification and recrystallization of metals, it has often been applied before to similar systems.[29,35] The followed procedure was very similar to the described by Andersen *et al.*[42] but the model could only be applied to the growth in thickness $<D_c>$, since the diameters $<D_a>$ are compromised by the resolution of the experiment. The fitted values for the Avrami exponents suggest that the reaction mechanisms at the three higher temperatures (264, 292 and 338 °C) are essentially limited by the diffusion process, while at the lowest temperature (245 °C) phase-boundaries seem to play a role too. Mathematical details and figures are compiled in the relevant section of the ESI.

## 3.2. *Ex situ* study

### High-resolution powder X-ray diffraction

Seven *ex situ* experiments of different durations were performed at 243 °C. High-resolution PXRD data were collected for these samples. Figure 7 shows the data measured for the sample reacted for 2 min in the spiral reactor, as well as the Rietveld models calculated for the individual phases.

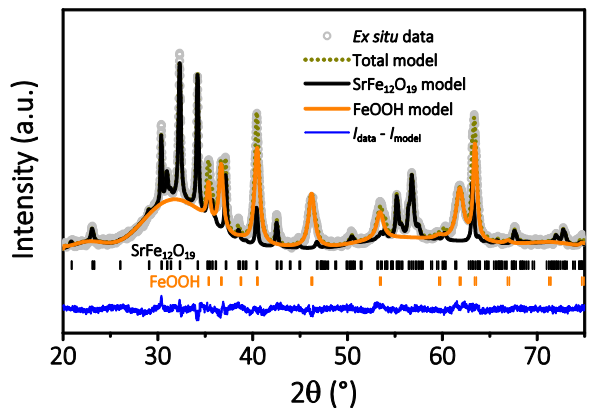

Figure 7. *Ex situ* PXRD experimental data for the 2 min long reaction in the spiral reactor, along with the corresponding Rietveld model. The measured data are represented by the grey circles, while the models for each of the phases are solid lines: black for $SrFe_{12}O_{19}$ and orange for FeOOH. The total model is represented by the green dotted line. The blue line at the bottom is the difference between the measured data and the calculated Rietveld model.

Figure 8(a) shows the time evolution of the refined weight fractions. Supporting the trend established in the *in situ* study, FeOOH is found to form initially to then gradually turn into $SrFe_{12}O_{19}$, following the same symmetric conversion observed for the *in situ* data. FeOOH is the only crystalline phase present at the shortest reaction times (20 s, 40 s) while pure $SrFe_{12}O_{19}$ is obtained for the longest reaction times (10 min, 20 min). Coexistence of the two phases is observed at the intermediate reaction times.

Crystallite sizes for the two phases are plotted in Figure 8(b) as a function of synthesis time. For FeOOH, the sizes along both crystallographic directions $<D_a>$ and $<D_c>$ maintain a more or less constant value of 17 nm and 6 nm, respectively. On the other hand, a significant growth takes place for $SrFe_{12}O_{19}$ within the first 5 min. Beyond that point, FeOOH disappears completely and the $SrFe_{12}O_{19}$ size remains more or less steady at around 180-200 nm and 60 nm for $<D_a>$ and $<D_c>$, respectively. The overall size-trend with reaction time in these *ex situ* studies is the same as for the *in situ* experiments, although the absolute values differ.

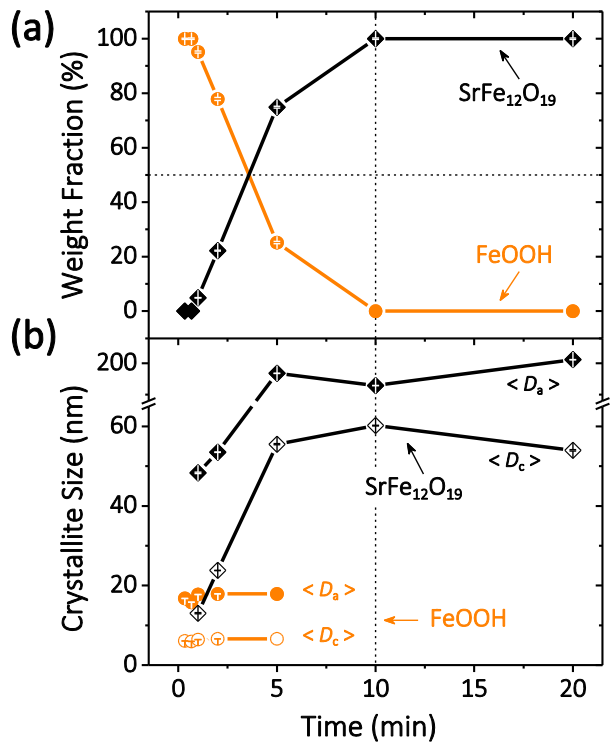

Figure 8. Results obtained from Rietveld refinement of PXRD data corresponding to the seven *ex situ* experiments. $SrFe_{12}O_{19}$ is shown in black and FeOOH in orange (a) Phase composition of the samples. (b) Volume-averaged crystallite sizes for both phases. Closed symbols show the sizes in the *a/b*-plane $<D_a>$, while sizes along the *c*-axis $<D_c>$ are represented by open symbols.

Transmission electron microscopy

TEM images were collected on selected *ex situ* synthesized samples: 20 s, 2 min and 10 min, attempting to capture the different stages of the reaction. Representative images for each sample are shown in Figure 9.

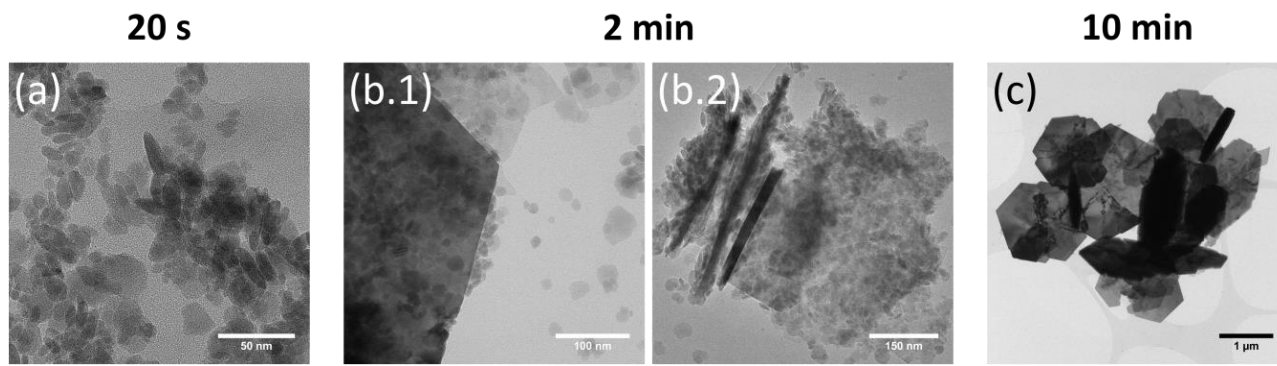

Figure 9. TEM images of *ex situ* samples, hydrothermally synthesized at 243 °C using the spiral reactor, with varying reaction times: (a) 20 s, (b.1, b.2) 2 min and (c) 10 min. Two different particle morphologies are discernible at the intermediate stages of the reaction (2 min). Hexagonal plate-like morphology of $SrFe_{12}O_{19}$ is confirmed.

PXRD confirms that FeOOH is the only crystalline phase present at the very beginning of the reaction. Figure 9(a) shows an image collected for the sample reacted for 20 s where disc-shaped nanoparticles are predominantly found. The long dimension of these discs is estimated to be between 10 and 30 nm, while the $<D_a>$ refined for this sample was 16.8(2) nm. This suggests monocrystallinity of these small FeOOH nanoparticles.

After 2 min of reaction, two very different particle morphologies are clearly discernible (see Figure 9(b.1) and (b.2)). Small FeOOH particles like those described above for the short times coexist with much larger particles with rather sharp edges, which are assigned to $SrFe_{12}O_{19}$. This observation is in good agreement with the two crystalline phases refined for the PXRD. Figure 9(b.1) shows a corner of one of a $SrFe_{12}O_{19}$ hexagonal platelet as well as small FeOOH particles attached to the plate surface. Likewise, the darker long particles on the left side of Figure 9(b.2) are identified as platelets seen edgewise.

After 10 min of reaction there is no sign of the small FeOOH particles and only large $SrFe_{12}O_{19}$ hexagonal platelets are found. The large mismatch between the $SrFe_{12}O_{19}$ sizes determined from PXRD and TEM can be explained by the entity the technique is actually probing. While PXRD is probing coherently scattering crystalline domains, *i.e.* crystallites, the transmission electron micrographs show whole particles. The large $SrFe_{12}O_{19}$ hexagonal plate-like particles consists of a number of crystallite domains. This can be discerned from the cracks or imperfections observed in the large platelets.

Attempting to quantify the elemental composition of the two different populations of particles, spatially resolved energy dispersive X-ray spectroscopy (EDS) was performed using the previously mentioned TALOS in scanning transmission electron microscopy (STEM) mode. No definitive conclusions could be drawn, as amorphous Sr may be present on the surface of the FeOOH particles. Representative examples are shown in the ESI.

## Magnetic hysteresis

Figure 10(a) shows the magnetic hysteresis curves for the samples prepared *ex situ* using the spiral reactor. Mass magnetization is plotted as a function of the effective field $H_{eff}$, *i.e.*, the magnetic field corrected for self-demagnetization. Figure 10(b) shows the critical points defined by the curves, *i.e.*, $M_s$, $M_r$ and $H_c$, as a function of time, as well as the calculated $BH_{max}$. The relevant values are listed in Table 1.

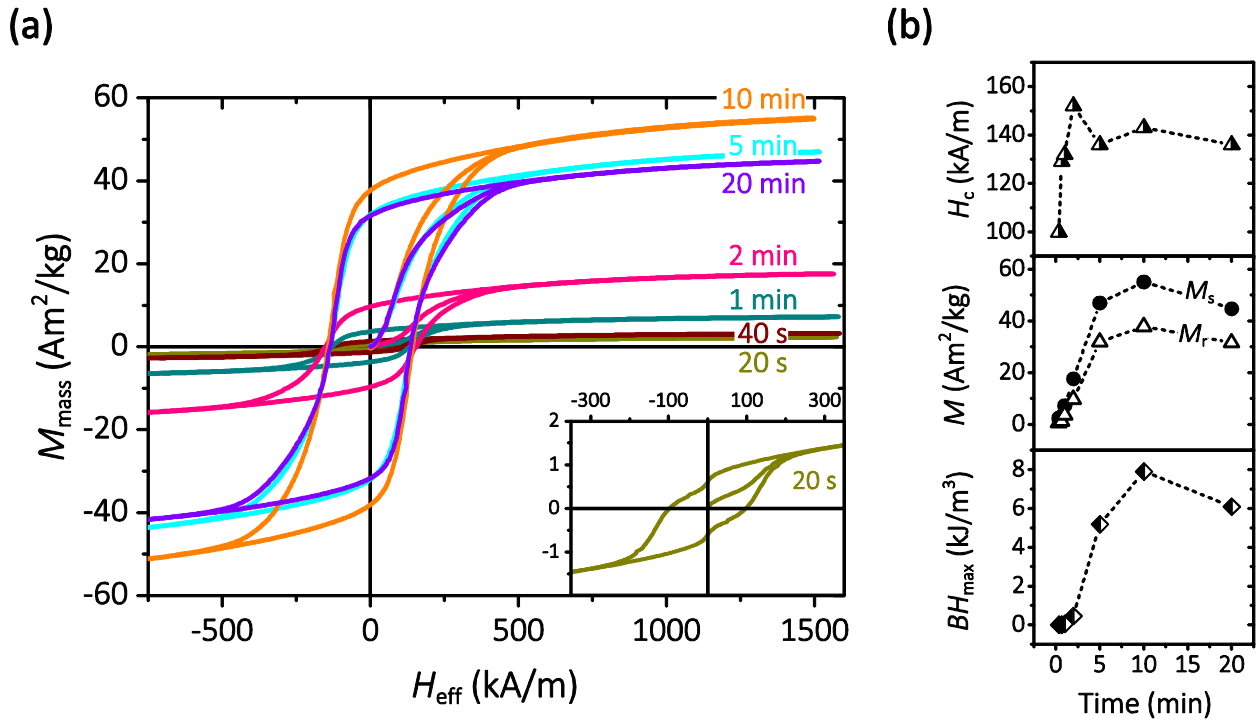

Figure 10. (a) Magnetic hysteresis curves measured at 300 K for all seven *ex situ* samples synthesized in the spiral reactor. The bottom right inset is a magnification of the hysteresis measured for the sample reacted for 20 seconds. (b) Magnetic parameters extracted from the curves vs. reaction time.

A general improvement of $M_s$, $M_r$, and $H_c$ with increasing reaction time is observed up to 5 min; at longer reaction times the values stay more or less constant. This trend is found to be in very good agreement with the crystallite thickness $<D_c>$ reported in Figure 8(b). The best-performing material of the series was obtained for a reaction time of 10 min. It presents a saturation $M_s$ of 55.1 Am²/kg, a remanence $M_r$ of 37.8 Am²/kg, a coercivity $H_c$ of 143 kA/m, and an energy product $BH_{max}$ of 7.90 kJ/m³.

Table 1. Time-dependent values of saturation mass magnetization ($M_s$), remanent mass magnetization ($M_r$), coercive field ($H_c$), and energy product ($BH_{max}$).

| reaction time | $M_s$ (Am$^2$/kg) | $M_r$ (Am$^2$/kg) | $H_c$ (kA/m) | $BH_{max}$ (kJ/m$^3$) |
|---|---|---|---|---|
| 20 s | 2.3 | 0.6 | 100 | $2.24 \times 10^{-4}$ |
| 40 s | 3.3 | 1.3 | 129 | $7.86 \times 10^{-3}$ |
| 1 min | 7.3 | 3.7 | 132 | $6.01 \times 10^{-2}$ |
| 2 min | 17.6 | 9.7 | 152 | $4.61 \times 10^{-1}$ |
| 5 min | 47.0 | 31.9 | 136 | 5.19 |
| 10 min | 55.1 | 37.8 | 143 | 7.90 |
| 20 min | 44.7 | 31.6 | 136 | 6.10 |

The bottom right inset in Figure 10(a) shows the hysteresis measured for the sample reacted for 20 s. This cycle is a two-step reversal hysteresis, which is usually a result of coexistence of two uncoupled magnetic phases in the measured sample. Under that assumption, the first of the species would be a ferro/ferrimagnetic (FM/FiM) phase with a $H_c$ of about 100 kA/m, which could easily correspond to hydrothermally synthesized $SrFe_{12}O_{19}$. The other magnetic species should then be an antiferromagnetic (AFM) or superparamagnetic phase with no coercivity, which would match the AFM behavior reported in literature for FeOOH.[55] Combining this observation with the PXRD data, we postulate coexistence of crystalline FeOOH with X-ray amorphous $SrFe_{12}O_{19}$ at the short reaction times, albeit further magnetic characterization would be required to confirm this hypothesis.

## 4. Discussion

This study confirms that it is thermodynamically favorable for $SrFe_{12}O_{19}$ to form thin hexagonal platelets with surface normal in the direction of the crystallographic c-axis when hydrothermally synthesized.[20] A plate-like morphology with the same crystallographic orientation is detected for FeOOH, although the hexagonal shape is less obvious from TEM. For FeOOH, a plate-like particle shape had been suggested before but, to the best of our knowledge, it had never been confirmed experimentally.[52]

A hydrothermal synthesis starting from a nitrate-based alkaline solution has been one of the most common methods used to synthesize $SrFe_{12}O_{19}$.[20,25] However, this is the first time FeOOH is detected as an intermediate in the process. This is interesting since we found the formation of FeOOH as an intermediate to be absolutely unavoidable for all the studied synthesis conditions. The discovery highlights the strength of an *in situ* experiment versus a regular laboratory synthesis. FeOOH eventually disappears to produce pure $SrFe_{12}O_{19}$ if reacted for long enough, which means one could easily miss it in a regular laboratory synthesis if the reaction time is long enough. The observation has consequences for the smallest size of $SrFe_{12}O_{19}$ that can be obtained phase pure through the described hydrothermal method. To further decrease the crystallite size of $SrFe_{12}O_{19}$ significantly high reaction rates are necessary.

Additionally, we believe that in previous studies with characterization based on microscopy, the six-line FeOOH could have been mistaken for tiny incipient $SrFe_{12}O_{19}$.[56] Distinguishing between these two phases may be challenging when the diffraction data are compromised by the high fluorescence background from iron, originated from the use of a Cu $K_\alpha$ source. Additionally, an insufficient peak resolution could also hide the presence of FeOOH, due to its similarity with $SrFe_{12}O_{19}$ in d- spacing and small crystallite size causing broad peaks in the diffraction pattern. The PXRD results presented in this work clearly highlight the need for including FeOOH in the Rietveld model in order to achieve a satisfactory description of the experimental data. A correct identification of the phases present in the sample is not only a matter of scientific rigor, but also a key point in the pursuit of understanding the relationship between magnetic properties and microstructure. Indeed, the occurrence of the six-line phase seems to have an effect on the magnetic properties of the produced $SrFe_{12}O_{19}$, as explained in the magnetic characterization section. Furthermore, the FeOOH content is a direct indication of how far along the reaction is. If FeOOH is still present, either higher temperatures or longer reaction times should be used in order to obtain single-phase $SrFe_{12}O_{19}$.

The *in situ* experiment at the lower temperature was replicated *ex situ* using the spiral reactor. The *ex situ* study corroborates the trends observed *in situ*. Weight fractions extracted from the *in situ* experiment at 245 °C hold good agreement with those obtained for the *ex situ* study (243 °C). In both cases single-phase $SrFe_{12}O_{19}$ is obtained after about 10 min of reaction. Crystallite sizes also show similar trends, although the *ex situ* values are significantly larger. This observation could have different origins: 1) the compromises made to the quality of the *in situ* data makes it difficult to extract absolute sizes.[39] 2) The post-synthesis treatment of the *ex situ* samples may cause an offset in size, *e.g.*, too low centrifugation speed would cause only the large nanoparticles to go to the bottom of the centrifuge tube, making it likely that smaller particles were discarded with the supernatant. 3) An insufficient instrumental resolution for the *in situ* measurements, *i.e.* the size broadening approaches the instrument resolution. Likewise is the *a/b*-size <$D_a$> for the *ex situ* long reaction times approaching the resolution limit of the high resolution powder diffraction, causing a large uncertainty on the absolute values of the sizes.

Unlike well-implemented autoclave reactors, the spiral reactor permits tuning of pressure, temperature and reaction time while still obtaining relatively large amounts of product. This allows for measurement of physical properties and, at the same time, replicating the fast heating rate of the *in situ* setup. Nevertheless, both the *in situ* and the spiral reactors could potentially be adapted to mimic the synthesis conditions in a conventional autoclave. This could be achieved by decreasing the heating rate and relying on internally built-up pressure instead of applying pressure externally like in the studies reported here. For the *in situ* setup, the hot-air gun could be programmed to follow a less steep temperature ramp. Slow heating experiments could also be performed with the spiral setup by immersing the reactor in the oil bath at a low temperature and subsequently heating it up, instead of preheating the oil to the set temperature. However, no control of the heating ramp is possible in this case.

Although scalability of chemical reactors is always a challenging task, it may be relatively easy in the case of the spiral reactor due to its design. The volume of the reactor could be increased simply by coiling the tubing more tightly and/or using a larger oil bath. This would not alter the heating rate, which would remain constant as long as the cross section of the tubing is maintained. Furthermore, the available temperature range could be extended to higher temperatures by substituting the oil bath with a fluidized sand bed.[57]

For the synthesis of phase pure $SrFe_{12}O_{19}$ it is necessary to use an excess of Sr compared to the stoichiometric Fe/Sr ratio of 12. This has been reported in various papers,[25,44,56] and we confirm this observation. The formation of FeOOH prior to $SrFe_{12}O_{19}$ could be part of the explanation for that excess of Sr needed. The ferrihydrite appears to be the kinetic product under these synthesis conditions, forming instantaneously as the high temperature is applied. Once the FeOOH is formed, it evolves into the most thermodynamically stable iron oxide, hematite, unless there is significant presence of $Sr^{2+}$ in the solution to form $SrFe_{12}O_{19}$. This is observed when preparing a precursor with the stoichiometric Fe/Sr molar ratio of 12 (data shown in the ESI). In addition, a recent experiment performed by Andersen et al. shows that a precursor solution containing only $Fe(NO_3)_3$ in an alkaline medium (in comparable concentrations and reacted in almost identical synthesis conditions) results in hematite formation directly, bypassing FeOOH.[43] Therefore, the presence of Sr appears to be a requirement for the initial formation of FeOOH in a strongly alkaline medium.

Regarding the magnetic properties of the obtained pure $SrFe_{12}O_{19}$, the $M_s$ is similar to values reported in the literature of as-synthesis powder obtained by hydrothermal synthesis without post calcination.[44,58,59] However, the $M_r$ of 37.8 $Am^2/kg$ measured for a reaction time of 10 min is relatively high for the moderate $H_c$ value. The remanence-to-saturation ratio $M_r/M_s = 0.69$ suggests that the platelets prefer to rest on their large surface, which causes some alignment. Park et al. reported similar $M_r$, but with 4 times larger $H_c$ compared to this work.[60] Despite the moderate $H_c$ achieved in this work, a significant enhancement is expected upon compaction of such plate-like powders into densified magnets. Saura-Múzquiz et al. recently reported an increase of approx. 250% in $H_c$ upon compaction of hydrothermally synthesized nanoplatelets by Spark Plasma Sintering (SPS).[25,26] Interestingly, the $H_c$ for 2 min has a larger value than expected within the series frame (152 kA/m). This discontinuity could be explained by the coexistence of $SrFe_{12}O_{19}$ with a secondary antiferromagnetic phase (six-line FeOOH). Large increases in $H_c$ have been reported when single-domain $SrFe_{12}O_{19}$ are embedded into an inorganic matrix.[61] This improvement is usually associated to a reduction of the inter-particle dipole-dipole interactions. It is suggested that something similar could be happening in this case, with the $SrFe_{12}O_{19}$ platelets surrounded by small FeOOH nanoparticles, as seen by the TEM images. Therefore, we reiterate on the importance of being aware of the presence of FeOOH.

## 5. Conclusions

Single-phase $SrFe_{12}O_{19}$ was synthesized through a hydrothermal method using two different experimental setups. Firstly, the synthesis was monitored in situ using synchrotron radiation. By doing so, the chemical process could be followed on a short time scale (time resolution = 5 seconds) and problems with post-synthesis treatment affecting observations were avoided. Small sample volumes and fast heating rates allowed nanoparticle formation in short reaction times (<10 min). The in situ studies, revealed the appearance of the six-line ferrihydrite prior to $SrFe_{12}O_{19}$ crystallization. The six-line FeOOH was invariably found as an intermediate phase in all the hydrothermal syntheses carried out, although a final conversion into $SrFe_{12}O_{19}$ was observed in all cases. The in situ studies also provided a handle to control the average crystallite size through time and reaction temperature, thereby controlling the magnetic properties. Larger crystallites were readily formed at the elevated temperatures, while smaller crystallites were obtained from synthesis at lower temperatures – although longer reaction times were needed to obtain single-phase $SrFe_{12}O_{19}$.

Reproduction of the *in situ* results at a larger scale was demonstrated through a purpose designed batch-type reactor. Almost instant heating rates (>10 °C/s) were likewise achieved using this novel reactor design, even with the relevant increase in volume. The *in situ* experiments were successfully reproduced using the new reactor. Higher quality PXRD data were collected on the *ex situ* samples, which combined with TEM images, allowed disclosing further details regarding the obtained product. Magnetic hysteresis curves were measured on the *ex situ* powders, and a correlation between the magnetic properties and the crystallite size could be established. Furthermore, the *ex situ* studies confirmed the appearance of the six-line ferrihydrite as a necessary intermediate in the reaction. This intermediate phase has been either ignored or mistaken for smaller $SrFe_{12}O_{19}$ crystals in previous works, due to the peak overlap between the two hexagonal structures. However, it is crucial to consider the presence of FeOOH in $SrFe_{12}O_{19}$ samples when interpreting the magnetic behavior of the material.

## Acknowledgments

The authors would like to thank financial support from the European Commission through the project NANOPYME FP7-NMP-2012-SMALL-6 NANOPYME (No. 310516). The work was likewise supported by the Danish National Research Foundation (Center for Materials Crystallography, DNRF93) and the Danish Research Council for Nature and Universe (Danscatt). Furthermore, support by the European Community's Seventh Framework Programme (FP7/2007-2013) CALIPSO under Grant Agreement No. 312284 is also acknowledged. Support is also gratefully accepted from the Danish Research Council for Technology and Production Sciences through a Sapere Aude grant (Improved Permanent Magnets through Nanostructuring).

# Graphic Table of Contents

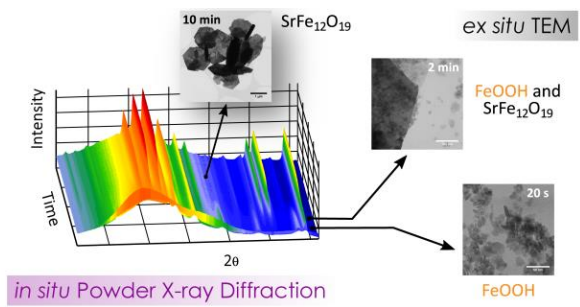

Hydrothermal synthesis of SrFe$_{12}$O$_{19}$ is followed *in situ* using PXRD, and successfully reproduced *ex situ* using a custom-designed batch-type reactor.